\documentclass[twocolumn, twocolappendix]{aastex63}

\usepackage{graphicx}	
\usepackage{color}
\usepackage{amsmath}	
\usepackage{amssymb}	
\usepackage{mathrsfs}
\usepackage{mathrsfs}

\newcommand{\msun}{M_\odot}

\newcommand{\zsun}{Z_\odot}

\newcommand{\msunyr}{M_\odot~{\rm yr}^{-1}}

\newcommand{\pc}{{\rm pc}}

\newcommand{\Muv}{M_{\rm UV}}
\newcommand{\SFR}{{\rm SFR}}

\newcommand{\K}{{\rm K}}
\newcommand{\beq}{\begin{equation}}
\newcommand{\eeq}{\end{equation}}

\shorttitle{Ultra-high Redshift Galaxies}
\shortauthors{Inayoshi et al.}

\begin{document}

\title{A Lower Bound of Star Formation Activity in Ultra-high Redshift Galaxies Detected with JWST:\\
Implications for Stellar Populations and Radiation Sources}

\correspondingauthor{Kohei Inayoshi}
\email{inayoshi@pku.edu.cn}

\author[0000-0001-9840-4959]{Kohei Inayoshi}
\affiliation{Kavli Institute for Astronomy and Astrophysics, Peking University, Beijing 100871, China}
\author[0000-0002-6047-430X]{Yuichi Harikane}
\affiliation{Institute for Cosmic Ray Research, The University of Tokyo, 5-1-5 Kashiwanoha, Kashiwa, Chiba 277-8582, Japan}
\author[0000-0001-9840-4959]{Akio K. Inoue}
\affiliation{Waseda Research Institute for Science and Engineering, Faculty of Science and Engineering, Waseda University, 3-4-1, Okubo, Shinjuku, Tokyo 169-8555, Japan}
\affiliation{Department of Physics, School of Advanced Science and Engineering, Faculty of Science and Engineering, Waseda University, 3-4-1, Okubo, Shinjuku, Tokyo
169-8555, Japan}
\author[0000-0002-1044-4081]{Wenxiu Li}
\affiliation{Kavli Institute for Astronomy and Astrophysics, Peking University, Beijing 100871, China}
\affiliation{Department of Astronomy, School of Physics, Peking University, Beijing 100871, China}
\author[0000-0001-9840-4959]{Luis C. Ho}
\affiliation{Kavli Institute for Astronomy and Astrophysics, Peking University, Beijing 100871, China}
\affiliation{Department of Astronomy, School of Physics, Peking University, Beijing 100871, China}

\begin{abstract}
Early results of {\it JWST} observations have delivered bright $z\gtrsim 10$ galaxy candidates in greater numbers than expected,
enabling construction of the rest-frame UV luminosity functions (LFs). 
The LFs contain key information on the galaxy assembly history, star formation activity, and stellar population
in the distant universe. 
Given an upper bound of the total baryonic mass inflow rate to galaxies from their parent halos estimated from abundance matching, 
we derive a lower bound on the product of the star formation and UV-photon production efficiency in galaxies at each redshift.
This stringent constraint requires a high efficiency ($\gtrsim 10-30\%$) converting gas into stars, 
assuming a normal stellar population with a Salpeter-like mass distribution.
The efficiency is substantially higher than those of typical nearby galaxies, but is consistent with those seen in starburst galaxies 
and super star clusters observed in the nearby universe.
Alternatively, the star formation efficiency may be as low as a few percent, which is the average value for the entire galaxy population at $z\simeq 6$, 
if the stellar population is metal-free and drawn from a top-heavy mass distribution that produces more intense UV radiation.
We discuss several other possible scenarios to achieve the constraint, for instance, energetic radiation produced from compact 
stellar-remnants and quasars, and propose ways to distinguish the scenarios by forthcoming observations.
\end{abstract}

\keywords{Galaxy formation (595); High-redshift galaxies (734); Quasars (1319); Supermassive black holes (1663)}

\section{Introduction}

The James Webb Space Telescope (JWST) is opening a new window into the most distant universe 
and unveiling the existence of the very first generation of galaxies at $z\gtrsim 10$.
Compiling the data of dropout galaxy candidates at high-redshifts found in the first JWST/NIRCam images,
the UV luminosity functions (LFs) at $z \sim 9-17$ are constructed \citep[e.g.,][]{Donnan_2022, Finkelstein_2022,Naidu_2022,Harikane_2022c}.
While the characteristic shapes of the LF determined with the JWST-identified galaxies nicely agree with those derived with Hubble Space Telescope,
the bright end is further extended to $\Muv< -23$ mag at $z\sim 10-13$ and $\Muv<-20$ mag at $z\sim 17$
without showing a sharp, exponential decline in the number density.
Moreover, the on-going JWST observations have already discovered such bright galaxies at $z\gtrsim 10$ more than 
expected based on previous estimates of their cosmic density.

In this {\it Letter}, we examine the hypothesis that the star formation rates (SFRs) in those $z\gtrsim 10$ galaxies have to
be lower (at most comparable) than the total baryonic mass inflow rates onto their dark matter (DM) halos.
Despite the simple constraint from mass continuity, we find a tension between the abundance of the observed bright galaxies 
and that of the brightest galaxies expected from the underlying halo mass function and its growth rate in the $\Lambda$CDM model.
The argument constrains the properties of star formation activity and population of those radiation sources,
suggesting (1) a substantially high star formation efficiency exceeding $\gtrsim 10-30\%$ for normal stellar populations,
(2) an extreme stellar population with a top-heavy initial mass function (IMF), (3) possible contributions
from accreting stellar-remnant black holes (BHs), and (4) quasars or super-Eddington accreting massive BHs.

Recently, \cite{Boylan-Kolchin_2022} pointed out a tension between the estimated galaxy stellar masses \citep{Labbe_2022} and 
those expected in $\Lambda$CDM-like models \citep[see also][]{Behroozi_Silk_2018}.
Our concept in this paper basically follows what they discussed, but in comparison with theoretical models we adopt UV luminosities,
which are more reliable than the stellar mass estimates.
We further extend the argument to the properties of stellar populations in those high-$z$ galaxies.
The tension regarding over-production of massive and bright galaxies could be also (partially) solved by assuming different stellar populations.
More tests of observations and theory are urgently needed as more new data of JWST are delivered.
\citep{Atek_2022, Castellano_2022, Donnan_2022, Finkelstein_2022, Labbe_2022, Morishita_2022, Naidu_2022,
Yan_2022, Harikane_2022c}.

\vspace{5mm}
\section{The upper bound of galaxy luminosity functions}\label{sec:method}

We construct the abundance of massive DM halos at $z\gtrsim 10$ from the Sheth-Tormen mass function \citep{Sheth_Tormen_2001},
adopting the initial linear power spectrum calculated with the Code for Anisotropies in the Microwave Background \citep[CAMB;][]{Lewis_2000}
and the growth factor for linear fluctuations \citep{Lukic_2007}.
Throughout this {\it Letter}, we assume a $\Lambda$CDM cosmology consistent with the latest constraints from Planck 
\citep{Planck_2020}.
For given halo mass $M_{\rm h}$ and redshift $z$, the mass growth rate of DM halos is characterized with an analytical function of the form
\citep{Fakhouri_2010}
\begin{align}
\dot{M}_{\rm h} \simeq & ~ 46.1~\msunyr \left(\frac{M_{\rm h}}{10^{12}~\msun}\right)^{1.1}\nonumber\\[5pt]
& \times (1+1.11z)\sqrt{\Omega_{\rm m}(1+z)^3+\Omega_\Lambda},
\label{eq:Mdot}
\end{align}
which is taken from the cosmological $N$-body simulations (Millennium simulations; \citealt{Boylan-Kolchin_2009}).
The $M_{\rm h}$- and $z$-dependence are consistent with those derived based on the extended Press-Schechter formalism 
\citep[e.g.,][]{Press_Schechter_1974,Bond_1991}, 
$d(\ln M_{\rm h})/dt \propto (1+z)^{5/2}$ \citep{Dekel_2013}.
The normalized value in Eq.~(\ref{eq:Mdot}) is the mean rate of the distribution of $\dot{M}_{\rm h}$, 
and is nearly twice higher than the median rate since the distribution has a positive tail.
Note that the cosmological parameters adopted in those simulations are out of date (e.g., $\sigma_8=0.9$).
\cite{Dong_2022} employed a set of high-resolution $N$-body simulations to study the merger rate of DM halos
and found that while the functional form is universal even with different combinations of cosmological parameters,
the rate normalization decreases by $\sim 0.1$ dex using up-to-date cosmological parameter sets 
that favor a smaller value of $\sigma_8=0.8102$ \citep{Planck_2020}.
They also show that other prescriptions of the halo growth rate \citep[e.g.,][]{Genel_2009,Stewart_2009} are consistent within the accuracy.
Therefore, the SFR and UV luminosity given below would be overestimated, and our model choice gives a conservative argument for our purpose.
The total baryonic inflow rate into a halo is given by $\dot{M}_{\rm b}=\dot{M}_{\rm h}(\Omega_{\rm b}/\Omega_{\rm m})$, where
the baryon fraction is $\Omega_{\rm b}/\Omega_{\rm m}=0.1621$ \citep{Planck_2020}.

Let us suppose that a fraction $f_\star$ of the gas accretes onto the galaxy and forms stars: $\SFR=f_\star \dot{M}_{\rm b}$.
The value of the conversion factor can be characterized with various effects and has been investigated based on the existing high-$z$ galaxy observations.
For high-$z$ galaxies at $10<z<15$, an empirical model for linking galaxy SFRs to the properties of host halo predicts
$f_\star \lesssim 0.01$ at $M_{\rm h}\sim 10^{9-10}~\msun$ \citep{Behroozi_2020}, which is the typical mass range
of the DM halos for bright, JWST-detected galaxies. 
A higher efficiency of $f_\star \sim 0.03-0.05$ is consistent with those inferred by abundance matching and the observed UV 
LF of galaxies at $z\sim 6$ \citep{Bouwens_2015} and by clustering analysis at $z\sim2-7$ 
\citep[][]{Harikane_2016, Harikane_2018,Harikane_2022a}.
In the local universe, more direct observations measuring the gas mass and SFR find that 
the global star formation efficiencies in galaxies are typically $f_\star \simeq 0.01-0.03$, but increase to $f_\star \gtrsim 0.3$ in starbust galaxies
\citep[e.g.,][]{Bigiel_2008,Kennicutt_2012}.
Moreover, super star clusters found predominately in starburst environments are expected to form from giant molecular clouds
in a short time at a high star formation efficiency of $f_\star \gtrsim 0.3-0.5$ to be gravitationally bound 
\citep[e.g.,][]{Lada_1984,Ho_1996,Keto_2005,Murray_2010}.
Recent numerical studies of star cluster formation from compact giant molecular clouds also suggest high efficiencies of $f_\star \simeq 0.2-0.3$ 
when an initial gas surface density is sufficiently high (\citealt{Kim_2018,Fukushima_2020,Fukushima_2021}; see also a review by \citealt{Krumholz_2019}).

\begin{table}
\renewcommand\thetable{1} 
\caption{UV luminosity-SFR conversion factor.}\vspace{-2mm}
\begin{center}
\begin{tabular}{cccc}
\hline
\hline
(1) & (2) & (3)& (4)\\
IMF & $Z$ & $\eta_{\rm UV}$ & $\epsilon_{\star, \rm UV}$\\
\hline 
Salpeter [0.1, 100]  &  0.02  &  $7.94\times 10^{27}$ & $2.79 \times 10^{-4}$  \\
Salpeter [0.1, 100]  &  0.0004 & $9.32\times 10^{27}$ &  $3.28 \times 10^{-4}$\\
Salpeter [50, 500]  &  0 & $3.57\times 10^{28}$& $1.26\times 10^{-3}$ \\
log-normal$^\dag$[1,500]  &  0 & $2.68\times 10^{28}$ & $9.42\times 10^{-4}$\\
\hline 
\end{tabular}
\label{tab:uv}
\end{center}
\tablecomments{Column 1: IMF mode with the mass range in $\msun$. 
Column 2: metallicity (the solar metallicity corresponds to $\zsun=0.02$). 
Column 3: the conversion factor from the SFR to the specific UV luminosity at $1500~{\rm \AA}$
in units of ${\rm erg}~{\rm s}^{-1}~{\rm Hz}^{-1}/(\msunyr)$. 
Column 4: the UV-radiation efficiency.
$\dag$ A log-normal distribution with a mean mass of $M_\star=10~\msun$ and dispersion of $\sigma_\star =1~\msun$ is assumed.
}
\end{table}

For a given SFR, we estimate the specific UV luminosity (in units of erg s$^{-1}$ Hz$^{-1}$) as
\begin{equation}
L_{\rm UV,\nu_0} = \eta_{\rm UV} \cdot \SFR,
\end{equation}
where the conversion factor $\eta_{\rm UV}$ is given by assuming the IMF and age of the stellar population \citep[e.g.,][]{Madau_Dickinson_2014}.
As a fiducial case, we adopt $\eta_{\rm UV}\simeq 7.94\times 10^{27}~{\rm erg}~{\rm s}^{-1}~{\rm Hz}^{-1}/(\msunyr)$
at $\lambda_0=1500$~\AA~($\nu_0\simeq 8.3$ eV), corresponding to a stellar population formed with a Salpeter IMF \citep{Salpeter_1955} 
in the mass range of $0.1-100~\msun$ with $Z=\zsun$.
The stellar age is set to $t_{\rm age}\gtrsim 100~{\rm Myr}$, where the photon production efficiency is saturated.

For convenience of the following discussion, we define the UV radiative efficiency of star formation for a given SFR as
\begin{equation}
\epsilon _{\star,\rm rad} \equiv \frac{L_{\rm UV}}{\SFR \cdot c^2} =2.79\times 10^{-4},
\end{equation}
where $L_{\rm UV}\equiv \nu_0L_{\rm UV,\nu_0}$.
As shown in Fig.~18 of \cite{Harikane_2022c}, the UV photon production efficiency increases by a factor of $\simeq 1.2$
as the metallicity decreases to $Z\simeq \zsun/50$.
More drastic enhancement of the efficiency is achievable in the limit of $Z\simeq 0$ \citep{Schaerer_2002,Schaerer_2003}.
For metal-free Population III stars (hereafter, Pop III), the efficiency becomes $3-4$ times higher when an extremely top-heavy IMF 
(a Salpeter IMF with $50-500~\msun$) and a moderately top-heavy IMF (log-normal with a characteristic mass of $M_\star = 10~\msun$, 
dispersion $\sigma_\star =1~\msun$, and wings extending from $1-500~\msun$; \citealt{Zackrisson_2011}). 
In Table 1, we list the values of the UV photon production efficiency and radiative efficiency for different IMF shapes.

\begin{figure*}
\begin{center}
{\includegraphics[width=170mm]{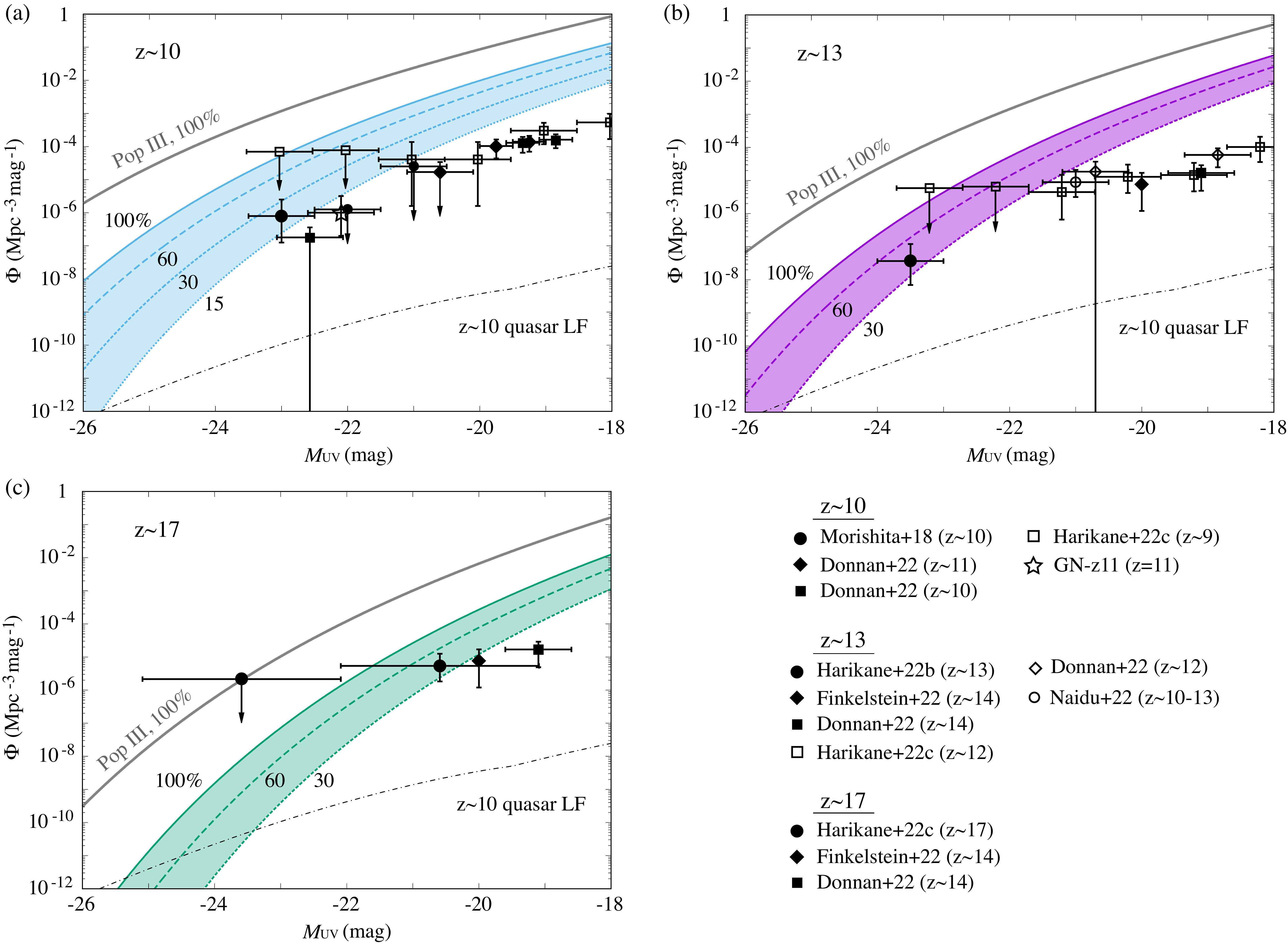}}
\caption{Galaxy LFs at three different redshift ranges of (a) $z\sim 10$, (b) $z\sim 13$, and (c) $z\sim 17$, 
along with the observed data points. The UV photon production efficiency per star formation rate is set to $\epsilon_{\star, \rm rad}=2.79 \times 10^{-4}$, 
assuming a stellar population with a Salpeter IMF in the mass range of $0.1-100~\msun$ with $Z=\zsun$. 
Each curve corresponds to the cases with different star formation efficiencies of $f_\star =1.0$ (solid), $0.6$ (long dashed), $0.3$ (short dashed), and $0.15$ (dotted). 
The grey thick curve indicates the {\it true} upper bound, where metal-free stellar population (i.e., Pop III stars) with
an extremely top-heavy IMF (the third model in Table 1) and $f_\star =1.0$ are considered.
The UV luminosity function data are taken from \cite{Oesch_2016, Morishita_2018, Donnan_2022, Harikane_2022b,
Harikane_2022c, Finkelstein_2022, Naidu_2022}.
For comparison, we overlay the quasar LF at $z\sim 10$ obtained a semi-analytical calculation (dashed-dotted; Li et al., in prep).
}
\label{fig:LF}
\end{center}
\end{figure*}

\vspace{2mm}
\section{Results}\label{sec:result}

In Fig.~\ref{fig:LF}, we show the galaxy LFs at three different redshift ranges of (a) $z\sim 10$, (b) $z\sim 13$, and (c) $z\sim 17$,
along with the observed data points \citep{Oesch_2016, Morishita_2018, Donnan_2022, Harikane_2022b,
Harikane_2022c, Finkelstein_2022, Naidu_2022}.
The UV photon production efficiency per star formation rate is set to $\epsilon_{\star, \rm rad}=2.79\times 10^{-4}$,
assuming a stellar population with a Salpeter IMF in the mass range of $0.1-100~\msun$ with $Z=\zsun$.
Each curve corresponds to the case with different star formation efficiency of $f_\star = 1.0$ (solid), $0.6$ (long dashed), $0.3$ (short dashed), and $0.15$ (dotted).
The region above $f_\star=1.0$ is prohibited for the given IMF because the SFR exceeds its strict upper bound, namely, 
the total baryonic mass inflow rate.

The LFs constructed with the procedure described in Section~\ref{sec:method} provide an upper bound of the UV luminosity 
for a given number density of galaxies at each redshift.
Overall, the brightest data point in each panel is closer to the shaded region and puts a stringent constraint on the star formation efficiency;
namely, $f_\star \gtrsim 15\%$ ($z\sim 10$) and $f_\star \gtrsim 30\%$ ($z\sim 13-17$).
The required efficiencies are substantially higher than the galaxy-scale average value for typical galaxies ($f_\star \sim {\rm a~few}~\%$),
but seem consistent with those seen in starburst galaxies and super star clusters ($f_\star \sim {\rm a~few~tens}~\%$).
In fact, this interpretation appears reasonable because the gas depletion timescale of $\sim O(100)$ Myr in nearby counterpart starburst environments is 
comparable to or a substantial fraction of the cosmic age at the redshift of interest ($t_{\rm H}\simeq 200-400$ Myr at $12<z<17$). 
Recently, \cite{Mason_2022} pointed out that the apparent overproduction of bright galaxies at $z\gtrsim 10$ would be caused by
extremely high SFRs with young stellar ages of $\sim 10-100$ Myr.

An alternative way to ease this stringent constraint is to consider a stellar population with a top-heavy IMF that yields a higher UV production efficiency.
If the IMF assumption is relaxed, the constraint is rewritten as 
\begin{equation}
f_\star \cdot \epsilon_{\star, \rm rad}\gtrsim 
\left\{
\begin{array}{l}
 4.05\\
 8.10
\end{array}
\right\} \times 10^{-5},
\label{eq:feta}
\end{equation}
where the values in the bracket are for $z\sim 10$ and $z\sim 13-17$, respectively.
In Fig.~\ref{fig:fstaIMF}, we show the parameter regions of $\epsilon_{\star, \rm rad}$ and $f_\star$
where the observed high-$z$ UV LFs (and the brightest data points) are consistently explained.
Adopting the most extreme stellar population with $Z=0$ that follows a Salpeter IMF with $50-500~\msun$, 
the UV production efficiency becomes as high as $ \epsilon_{\star, \rm rad}\simeq 1.26\times 10^{-3}$.
In the extreme case, the star formation efficiency is reduced to $f_\star \gtrsim 0.03-0.07$, which is moderate 
and consistent with the canonical value of $f_\star \simeq 0.03-0.05$ for the global efficiency observed in local, star-forming galaxies 
\citep[e.g.,][]{Bigiel_2008,Kennicutt_2012}.
In Fig.~\ref{fig:LF}, we also overlay the true upper bound for the UV LF at each redshift (grey thick curve), adopting 
Pop III stars with the top-heavy IMF and the 100\% star formation efficiency ($f_\star=1.0$).
All the observation data points are consistent with the $\Lambda$CDM model, but the upper limit of the abundance for 
the brightest sources at $z\sim 17$ is close to this strict upper bound.

Recent numerical simulations suggest that the stellar IMF in lower-metallicity environments tends to be top-heavy 
and the star formation efficiency is as high as $\sim 50\%$ in extremely metal-poor clouds with $Z\lesssim 10^{-3}~\zsun$ \citep{Chon_Omukai_2021}.
On the other hand, in moderate low-metallicity environments with $10^{-2}<Z/\zsun<1$, the efficiency shows a positive correlation
with the metallicity because line cooling via C~{\sc ii} and O~{\sc i} suppresses expansion of ionized regions
and regulates stellar feedback, leading to a high star formation efficiency \citep{Fukushima_2021}.
As an example, Fig.~\ref{fig:fstaIMF} shows the star formation efficiencies obtained in \cite{Fukushima_2021},
where a Chabrier IMF \citep{Chabrier_2003} is adopted.
We note that a Chabrier IMF yields an $\sim 1.6$ times higher $\epsilon_{\star, \rm rad}$ than that for a Salpeter IMF 
with the same mass range and metallicity \citep{Madau_Dickinson_2014}, and thus reduces the required star formation efficiency to $f_\star \gtrsim 9-19\%$ (see Eq.~\ref{eq:feta}).
The star formation efficiency depends on various parameters characterizing the cloud initial conditions, 
but sensitively on the gas surface density of the parent cloud.
Combined with numerical results for a wider range of metallicities and cloud initial conditions, we need to further explore the possible parameter 
space on Fig.~\ref{fig:fstaIMF} where the abundance of UV luminous galaxies at $z\gtrsim 10$
can be consistently explained.

We also note that the conditions for $f_\star \epsilon_{\star, \rm rad}$ in Eq.~(\ref{eq:feta}) would still be a conservative
lower bound because we neglect dust attenuation, which decreases the UV luminosities. 
Recently, \cite{Ferrara_2022} claimed that the weak evolution of the bright end of the LF at $7<z<14$ 
can be explained with negligible dust attenuation.
The ALMA observations of GHZ2/GLASS-z13 \citep{Castellano_2022,Donnan_2022,Naidu_2022}, 
one of the brightest and most robust $z>10$ galaxy candidates identified with JWST, have reported a non-detection of dust continuum
\citep{Bakx_2022,Popping_2022}.
This result suggests low metallicity and negligible dust attenuation, consistent with the blue UV slope observed by JWST.
However, there are also several lines of evidence that heavily obscured galaxies do exist at $z>6-7$, as reported by
ALMA \citep{Fudamoto_2021,Inami_2022} and JWST observations \citep{Rodighiero_2022}.
Further follow-up observations with ALMA for JWST-identified $z>10$ galaxies would shed light on the physics regarding dust formation and ejection
during the very first galaxy assembly.

\vspace{2mm}
\section{Discussion}

\begin{figure}
\begin{center}
{\includegraphics[width=85mm]{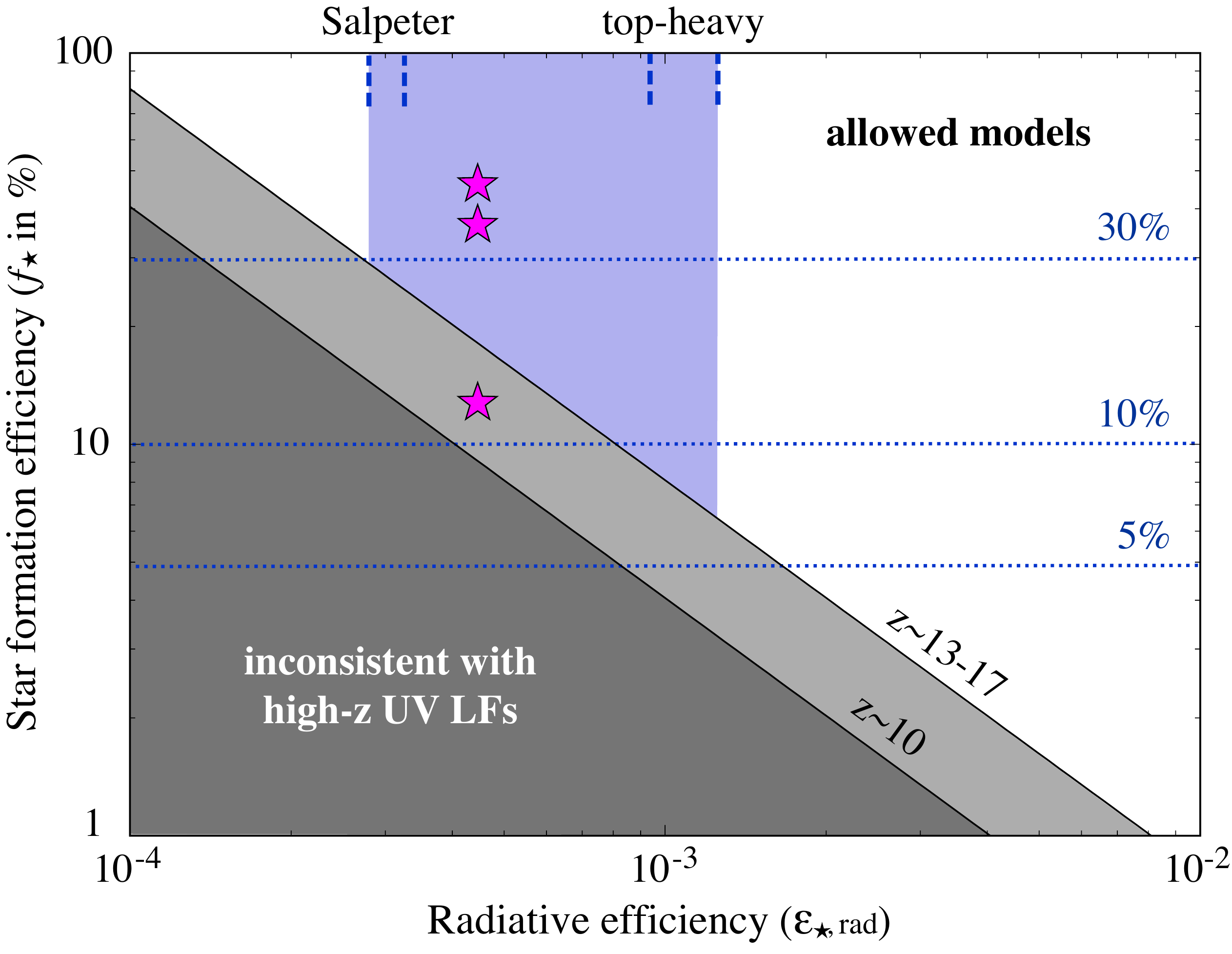}}
\caption{Parameter regions of the UV radiative efficiency ($\epsilon_{\star, \rm rad}$) and star formation efficiency ($f_\star$)
where the observed high-$z$ UV LFs (and the brightest data points) are consistently explained.
The blue vertical tickmarks indicate the values of $\epsilon_{\star, \rm rad}$ for the Salpeter and top-heavy IMFs
discussed in the text (see also Table 1), and horizontal lines show a constant value of $f_\star$.
For reference, the star formation efficiencies obtained from numerical simulations by \cite{Fukushima_2021} are shown with star symbols
($f_\star \simeq 0.13$, $0.37$, and $0.47$ for an initial gas surface density of $\Sigma_{\rm cl}=320$, $800$, and $1300~\msun~\pc^{-2}$, respectively).
Note that since the simulations adopted a Chabrier IMF, the corresponding value of $\epsilon_{\star,\rm rad}$ is approximately 1.6 times
higher than that for a Salpeter IMF with the same mass range and metallicity \citep{Madau_Dickinson_2014}.
}
\label{fig:fstaIMF}
\end{center}
\end{figure}

As discussed in Section~\ref{sec:result} (see also Figs.~\ref{fig:LF} and \ref{fig:fstaIMF}), the abundance matching procedure yields a stringent lower bound 
on the product value of $f_\star \epsilon_{\star, \rm rad}$ at redshifts of $z\sim 10-17$.
To achieve the requirement, we discuss two canonical interpretations \citep[see also][]{Harikane_2022c}: \vspace{2mm}\\
1. Extremely efficient episodes of metal-poor star formation with $f_\star \gtrsim 0.15-0.3$ and $Z\lesssim 1.0-0.1~\zsun$.
Note that the high efficiency is consistent with those seen in nearby starburst galaxies and super star cluster whose gas depletion timescale is
comparable to the cosmic age at $12<z<17$. \vspace{2mm}\\
2. Relatively efficient episodes of metal-free star formation with $f_\star \gtrsim 0.03-0.07$ and an extremely top-heavy IMF
with mean mass of $\langle M_\star \rangle \gg 10~\msun$. \vspace{2mm}\\
The parameter space of $\epsilon_{\star, \rm rad}$ and $f_\star$ in between the two cases are also allowed (see the blue shaded region in Fig.~\ref{fig:fstaIMF}).
In addition to these two options, we propose two other possible solutions:\vspace{2mm}\\
3. 
With a top-heavy IMF with $\langle M_\star \rangle \gtrsim 10~\msun$, 
a large fraction of those massive stars ends up in BHs via gravitational collapse.
For instance, for a Salpeter IMF in the mass range of $1-100~\msun$ or a top-heavy IMF we considered above, 
the mass fraction of massive stars forming BHs in a given mass budget is estimated as $f_\bullet \simeq 0.2$ and $\simeq 1.0$, respectively.
Here, non-rotating stars of zero-age main-sequence mass $\geq 20~\msun$ are assumed to leave remnant BHs. 
With those values, the total UV luminosity owing to accreting BHs is estimated as
\begin{align}
\frac{L_{\bullet, \rm UV}}{\SFR~ c^2} &= f_\bullet~\epsilon_{\rm \bullet,rad} ~f_{\rm UV,X},\\
& \simeq 1.0\times 10^{-3}~ f_{\bullet} \left(\frac{\epsilon_{\rm \bullet,rad}}{0.1}\right) \left(\frac{f_{\rm UV,X}}{0.01}\right).\nonumber
\label{eq:eta_BH}
\end{align}
The radiative efficiency of an accreting BH ranges from $\epsilon_{\rm \bullet,rad}\simeq 0.056 - 0.42$ depending on the BH spin (co-rotating case).
The conversion factor from X-rays (approximately the bolometric one) to the UV band is set to $f_{\rm UV,X}=0.01$ for $M_\bullet=100~\msun$ 
because stellar mass BHs produce most radiation in the X-ray band \citep[e.g.,][]{Tanaka_2012}.
We note that the value of $f_{\rm UV,X}$ is higher for heavier BHs because the peak frequency in the spectrum is shifted to lower energy \citep[e.g.,][]{Kato_2008}.
Therefore, the contribution from accreting massive stellar remnants, which would be observed as ultra-luminous X-ray sources, is comparable to or smaller than 
those from their progenitor stars for a given stellar IMF.
\vspace{2mm}\\
4. Quasars can be brighter than galaxies due to efficient release of gravitational energy of accreting matter into radiation.
For instance, if we assume that the UV luminosities of $z\simeq 13$ galaxies are powered only by accreting BHs, 
the inferred BH mass could be $\sim 10^8~\msun$, adopting Eddington-limited accretion rates \citep{Pacucci_2022}.
However, recent theoretical and empirical models for quasar LFs suggest that the predicted quasar number density at $z\sim 13$ 
is 2 orders of magnitude lower than the observed abundance (e.g., Li et al. 2022; \citealt{Finkelstein_Bagley_2022}).
In addition, as discussed in \citet{Harikane_2022c}, most of the bright galaxy candidates at $z\gtrsim10$ identified with JWST show 
morphologies more extended than the point-spread function, indicating that these candidates are not point sources such as quasars.
If super-Eddington accretion is allowed, the BH mass required to explain the luminosity is lowered and the number density of
less massive BHs is higher \citep{Inayoshi_2022a}.
\vspace{3mm}

The first two scenarios are expected to be distinguishable by photometric color selection and 
emission-line diagnostics with follow-up spectroscopy because metal-free stellar populations with a top-heavy IMF 
yield a characteristic radiation spectrum in the rest-frame UV band, such as prominent Ly$\alpha$ and He~{\sc ii} $\lambda$1640 
\citep{Tumlinson_Shull_2000,Schaerer_2002, Schaerer_2003, Inoue_2011,Zackrisson_2011}.
In the third scenario, a large number of stellar-mass BHs emit X-rays with a total luminosity of $L_{\rm X}\sim 10^{46}~{\rm erg}~{\rm s}^{-1}$.
This level of intense X-ray emission will be detectable with the planned Lynx satellite\footnote{https://wwwastro.msfc.nasa.gov/lynx/docs/science/blackholes.html} 
even at $z>10$.
The fourth scenario can be also tested with quasar selection criteria (photometric colors and characteristic nebular emission lines).
Several different methods for selecting accreting seed BHs have been proposed \citep{Inayoshi_2022b, Nakajima_2022, Goulding_Greene_2022}.
In addition, the morphological properties of galaxies are thought to be useful to differentiate quasars from galaxies. 
Although most of the bright $z\gtrsim 10$ galaxy candidates are extended sources, the galaxy size tends to decrease as $(1+z)^s$, 
where $s\simeq -1.19$ \citep{Ono_2022}.
In fact, a bright galaxy at $z \sim 12$ in their sample has an extremely compact size that is extended compared to the point-spread function,
and its surface brightness is fit by quasar+galaxy composite profiles. 
Thus, more detailed analysis is required for the morphological diagnostics for ultra-high redshift galaxies.
The third and fourth options are distinguishable from different X-ray spectra (a collection of high-mass X-ray binaries versus quasars).
Moreover, while quasars likely show some X-ray variability associated with accretion processes in the nuclei, 
variation of individual stellar-mass BHs would be diluted owing to their incoherent, non-synchronized variability.
Thus, observing X-ray time variability enables another independent diagnostic of the radiation source\footnote{
UV time variability would help distinguish the quasar scenario from the others. However, the time scale is expected to be 
the order of $\gtrsim 10~{\rm days}~[(1+z)/11]$ in the observer's frame, which is is substantially longer than that seen in the X-ray band.}.

Throughout this {\it Letter}, we adopt UV luminosities in comparison with theoretical upper limits of the galaxy number density.
This is because UV luminosities are more reliable than stellar mass estimates.
For instance, a galaxy at $z\simeq 17$, first reported by \cite{Donnan_2022} (CEERS 93316; $M_\star \simeq 10^9~\msun$ at $z = 16.6-16.7$)
was confirmed by \cite{Harikane_2022c} to have a mass consistent with the $\Lambda$CDM model.
Although \cite{Naidu_2022b} has subsequently claimed that this object is ``overmassive" (they report $M_\star \sim 5\times 10^9~\msun$),
we find that the abundance of bright high-$z$ galaxy candidates identified with JWST is still consistent
with star formation in $\Lambda$CDM cosmology \citep[see also][]{Mason_2022}, but places intriguing constraints on 
the stellar population and radiation sources.
The discrepancy arises from the difference in constraints from an integral quantity (stellar mass) and a differential quantity
(SFR and UV luminosities), as well as the uncertainty of galaxy mass estimates.

Although there are several observational uncertainties, most of the galaxies at $z\gtrsim10$ discussed in this paper are candidates robust against these uncertainties, 
especially bright ones that indicate high star formation or UV-radiation efficiency.
These bright candidates show clear breaks and blue colors redward of the break, consistent with $z\gtrsim10$ galaxies, and are identified in multiple studies as high-redshift galaxies \citep{Naidu_2022,Castellano_2022,Donnan_2022,Finkelstein_2022,Harikane_2022c}.
This indicates that differences in the photometry and in spectral fitting assumptions do not affect the selection process of those galaxies.
Indeed, \citet{Harikane_2022c} use stricter criteria than other studies (e.g., $\Delta \chi^2>9$, see their Section 3.3) to remove low-redshift interlopers, and they still obtain 
a high number density of UV-luminous candidates at $z\sim12-17$ (see Fig.~\ref{fig:LF}).
One caveat in the early JWST datasets is a zeropoint offset in the NIRCam data, which is due to a mismatch between pre-flight and in-flight reference files and detector-to-detector variations.
However, the resulting zeropoint offset is estimated to be small, less than $\sim20\%$ \citep{Rigby_2022}, which does not affect the robustness of these bright galaxy candidates.
The offset may change the UV luminosity estimate of these galaxies by up to $\sim20\%$, which is not sufficient to change the conclusions of this paper.

Of course, spectroscopy is mandatory to confirm the existence of these bright galaxy candidates.
In addition to confirming the redshifts, spectroscopic data will also allow us to investigate the physical origin of these bright candidates by examining the strengths of emission lines such as He~{\sc ii} $\lambda$1640 and C~{\sc iv} $\lambda$1549.
The currently used JWST data constitute about $10\%$ of the entire survey planned in Cycle 1.
Near-future photometric and spectroscopic observations with JWST will significantly improve our understanding of the early universe beyond $z\sim10$.

\acknowledgments
We greatly thank Hajime Fukushima, Jia Liu, Kazuyuki Omukai, Masafusa Onoue, Masami Ouchi, Shun Saito, and Kazuyuki Sugimura for constructive discussions. 
We acknowledge support from the National Natural Science Foundation of China (12073003, 12003003, 11721303, 11991052, 11950410493), 
and the China Manned Space Project (CMS-CSST-2021-A04 and CMS-CSST-2021-A06). 
A. K. I. acknowledge support from NAOJ ALMA Scientific Research Grant Code 2020-16B.
Y. H. is supported by the joint research program of the Institute for Cosmic Ray Research (ICRR), University of Tokyo, and
JSPS KAKENHI grant No. 21K13953.

\bibliography{ref}{}
\bibliographystyle{aasjournal}


\end{document}